\documentclass[prl,twocolumn,showpacs,preprintnumbers,amsmath,amssymb]{revtex4}

\usepackage[dvips]{graphics}
\usepackage{amsmath, amsthm, amssymb}
\usepackage{graphicx}
\usepackage{dcolumn}
\usepackage{bm}

\newcommand{\bra}[1]{\langle #1|}
\newcommand{\ket}[1]{|#1\rangle}

\newtheorem{theorem}{Theorem}
\newtheorem{definition}{Definition}
\newtheorem{lemma}{Lemma}

\usepackage{graphicx}
\usepackage{dcolumn}
\usepackage{bm}

\begin{document}

\title{Are all maximally entangled states pure?}

\author{D. Cavalcanti}\email{dcs@fisica.ufmg.br}
\affiliation{Departamento de F\'{\i}sica - CP 702 - Universidade
Federal de Minas Gerais - 30123-970 - Belo Horizonte - MG -
Brazil}
\author{F.G.S.L. Brand\~ao}\email{fgslb@ufmg.br}
\affiliation{Departamento de F\'{\i}sica - CP 702 - Universidade
Federal de Minas Gerais - 30123-970 - Belo Horizonte - MG -
Brazil}
\author{M.O. \surname{Terra Cunha}}\email{tcunha@mat.ufmg.br}
\affiliation{Departamento de Matem\'atica - CP 702 - Universidade
Federal de Minas Gerais - 30123-970 - Belo Horizonte - MG -
Brazil}

\begin{abstract}
We study if all maximally entangled states are pure through several
entanglement monotones. In the bipartite case, we find that the same
conditions which lead to the uniqueness of the entropy of
entanglement as a measure of entanglement, exclude the existence of
maximally mixed entangled states. In the multipartite scenario, our
conclusions allow us to generalize the idea of monogamy of
entanglement: we establish the \textit{polygamy of entanglement},
expressing that if a general state is maximally entangled with
respect to some kind of multipartite entanglement, then it is
necessarily factorized of any other system.
\end{abstract}

 \maketitle

%Arising as a counter intuitive phenomenon, entanglement has become the background
%for warm debates since it was noted \cite{1}. Understanding the non-classical correlations described by entanglement is a question that is in the heart of the fundamentals of quantum theory. In the last decades a new field raised: quantum information. For it, entanglement is a resource, yet to be completely understood \cite{2}.

One of the most striking differences between classical and quantum
correlations is the restricted capability of quantum states to share
entanglement. This so called \textit{monogamy of entanglement} has
been increasingly studied in the last years \cite{CKW}, and is
related to the security of quantum cryptographic protocols based on
entanglement (it limits the amount of correlations which an
eavesdropper can have with the honest parties). %The CKW inequality,
%established firstly
%or systems of qubits
%in Ref. \cite{CKW} and extended
% general systems
%in Ref. \cite{OF}, is a quantitative statement about the trade of
%entanglement in multi-partite states. Another aspects of the
%monogamous character of entanglement has been also linked with
%classical correlations \cite{KW}.
%was considered by Koashi and Winter
%, who shown that the amount of entanglement between two quantum
%systems restricts not only the entanglement, but also classical
%correlations between those and other systems.

The discussion about the monogamy of entanglement usually begins
with the apparent straightforward fact that maximally entangled
states are pure. This means that when two systems are as much
entangled with each other as it is possible, they cannot be
entangled and even classically correlated with any one else. In
this paper we analyze the trustiness of this ``common sense''
under the view of several entanglement measures. On one hand, it
is shown that it is not true for general entanglement monotones,
failing at least in the best separable approximation measure
\cite{KL} and in the indicator measure \cite{VP}. On the other hand,
we prove that for the majority of entanglement quantifiers
maximally entangled states are indeed pure. In particular, we
consider the quantifiers related to entanglement witnesses and, in
special, the generalized robustness of entanglement. With the help
of the \textit{witnessed entanglement} \cite{BV1}, we introduce the
idea of polygamy of entanglement, which states that \emph{if a
multipartite state is maximally entangled with respect to a given
kind of multipartite entanglement, then it must be pure}
\footnote{Note that we use the term \emph{polygamy} in
the sense of a marriage among multiple partners. This word has
appeared before, in Ref. \cite{AI04}, in the context of symmetric
multipartite Gaussian states, as opposing to monogamy, in the
sense that states which maximizes a certain pair-entanglement
quantifier can also maximize the multipartite version of it.} .

In order to avoid future confusion it is important to stress
that the idea of \emph{mixed maximally entangled states (MMES)}
 discussed here is, although related, different from the idea of
 \emph{maximally entangled mixed states (MEMS)} presented in the Refs. \cite{VAM}
 (this is the reason for the exchange of words). In their articles the authors
 address the following question: what is the highest value of entanglement that
  states with a given purity
(mixing) can present? In our work we study
if, given the maximum value of entanglement,
there is some mixed state which reaches it.

Before proceeding to show the main result of this Letter, we
present two simple results, albeit important, valid for every
convex quantifier.

\begin{theorem}\label{teo1}
According to all convex entanglement quantifiers there is at least
one maximally entangled pure state.
\end{theorem}
\begin{proof} Given a density operator $\rho=\sum_i{p_i\ket{\psi_i}\bra{\psi_i}}$
 and a convex quantity $E$, it holds $E(\rho) \leq \sum_{i}p_{i} E(\ket{\psi_{i}}\bra{\psi_{i}})$,
 for every ensemble decomposition ${\cal f}p_{i}, \psi_{i}{\cal g}$ of $\rho$. Thus, we see
 that there must be a $\ket{\psi_i}$ such that $E(\ket{\psi_i}\bra{\psi_i}) \geq E(\rho)$.
\end{proof}

Moreover, from the convex condition it is easily seen that for a
mixed state to be maximally entangled (with respect to $E$), there
must be an ensemble description with all $\ket{\psi_i}$ maximally
entangled.
\begin{theorem}\label{teo2}
If $\rho$ is a mixed maximally entangled state with respect to the
convex measure $E$, then all states in the subspace spanned by the
eigenvectors of $\rho$ are maximally entangled.
\end{theorem}

\begin{proof}
According to the unitary freedom in the ensemble for density matrices
theorem \cite{NC},
the sets $\{ p_i, \ket{\psi_i} \}$ and $\{ q_j,
\ket{\phi_j} \}$ generate $\rho$, i.e.
\begin{equation}
\rho = \sum_{i}p_{i}\ket{\psi_i}\bra{\psi_i} =
\sum_{j}q_{j}\ket{\phi_j}\bra{\phi_j},
\end{equation}
if and only if $\sqrt{p_i}\ket{\psi_i}=\sum_j u_{ij}\sqrt{q_j}\ket{\phi_j}$,
with $ \{ \ket{\psi_i} \}$ and $\{ \ket{\phi_j} \} $ being
normalized vectors, $u_{ij}$ a complex unitary matrix and one can
`pad' whichever set of vectors $\sqrt{p_i} \ket{\psi_i}$ or
$\sqrt{q_j} \ket{\phi_j}$ is smaller with additional null vectors
 so that the two sets have the same number of elements.

Since each pure state term in any convex decomposition of $\rho$ must
be
maximally entangled, we find that the state $\ket{\psi_i}=\sum_j c_{ij} \ket{\phi_j}$, with coefficients
$c_{ij} = u_{ij}\sqrt{\frac{q_j}{p_i}}$, must be
maximally entangled as well. The result follows letting ${\cal
f}\ket{\phi_j}{\cal g}$ be the eigenvectors of $\rho$
and noting
that for a fixed $i$, the vector $c_{ij}$ can have arbitrary
elements belonging to the hypersphere $\sum_{j}c_{ij}^{2} = 1$.
\end{proof}

It is possible to extend entanglement measures defined for pure
states to the whole state space with the convex-roof construction.
Given the quantity $E$, its convex-roof is
\begin{equation}\label{convroof}
E(\rho) = \min_{{\cal f}p_{i}, \psi_{i}{\cal g}}p_{i}E(\psi_{i}).
\end{equation}
It can be shown that $E(\rho)$ is an entanglement monotone
whenever $E(\psi)$ is. From Eq. \eqref{convroof} we see that, for
convex-roof based measures, Theorem \eqref{teo2} gives necessary
and \textit{sufficient} conditions for the existence of mixed
maximally entangled states.

We believe the existence of a $n$-dimensional subspace, with $n
\geq 2$, formed only by maximally entangled states is a very
demanding condition, so that for general convex entanglement
measures the maximally entangled states are pure. One might then
conjecture that this is true for all entanglement monotones
\cite{Vidal}. However, for the indicator measure \cite{VP} defined as
1 for entangled states and 0 for separable states, which is
obviously an entanglement monotone, every entangled state is
maximally entangled. Furthermore, using the result of Ref.
\cite{P} which for every $k$-partite Hilbert space $H$ there exists
an entangled subspace of dimension $d_{1}d_{2}...d_{k} - (d_{1} +
d_{2} + ... + d_{k}) + k - 1$, we find that also for the
convex-roof indicator measure maximally entangled states can be
mixed.

\begin{theorem}
According to every bipartite entanglement measure such that all its
maximally entangled pure states have maximum Schmidt rank possible
\footnote{The Schmidt rank of a bipartite pure state is the number
of non-null Schmidt coefficients in its Schmidt decomposition.},
there do not exist maximally entangled mixed states.
\end{theorem}
\begin{proof}
By theorem \eqref{teo2}, it must exist a subspace of maximally
entangled pure states. By hypothesis, they all have maximum Schmidt
rank. Take two of them,
\begin{equation}
\ket{\psi} = \sum_{ij}c_{ij}\ket{ij}, \hspace{0.3 cm} \ket{\phi} =
\sum_{ij}d_{ij}\ket{ij}.
\end{equation}
If we look at $c_{ij}$ and $d_{ij}$ as coefficients of square
matrices $C$ and $D$, maximum Schmidt rank is equivalent to
invertibility of the matrix. However, it always exists $\alpha$,
$\beta \in \mathcal{C}$ such that $\alpha C+\beta D$ is not
invertible (take $\frac{-\alpha}{\beta}$ as an eigenvalue of
$C^{-1}D$), and $\alpha\ket{\psi}+\beta\ket{\phi}$ does not have
maximum Schmidt rank.
\end{proof}
This theorem applies to a number of important and well-studied
entanglement measures. Consider first the entanglement of formation
\cite{BVSW} and the relative entropy of entanglement \cite{VP}. They
are both convex and equal to the entropy of entanglement ($E_{E}$)
in pure states. As all maximally entangled states of $E_{E}$ are
singlets (which have maximum Schmidt rank), it follows from theorem 3 that neither of them allows
MMES. The same argumentation is valid for the negativity \cite{VW}
and the concurrence \cite{WRC}. We can go even further and establish
the following result:
\begin{theorem}
For all asymptotic continuous and partially additive entanglement
monotones, all maximally entangled states are pure.
\end{theorem}
\begin{proof}
From the \textit{uniqueness theorem for entanglement measures}
\cite{HHH}, we have that every entanglement measure $E$ fulfilling
the conditions of the theorem obey $E_{D} \leq E \leq E_{F}$, where
$E_{D}$ and $E_{F}$ are the distillable entanglement and the
entanglement of formation respectively. Hence the result follows
straightforwardly from the fact that $E_{F}$ does not have MMES and
$E_{D}=E_{F}$ to pure states.
\end{proof}

The situation is much more subtle when we are dealing with
multi-partite entanglement. In this case we have to specify which
kind of entanglement we are talking about \cite{DC}. This is because
we could be interested in studying the entanglement among different
partitions of the whole system.
%For example, in a system composed of three parties $A$, $B$, and $C$,
%we can
%look for the entanglement between the partitions $A-BC$, $B-CA$,
%$C-AB$, or
%also for the genuine tripartite entanglement among $A-B-C$.
Furthermore in the multi-partite context, other relevant questions
arise in order to classify entangled states, as there are different
classes of inequivalent states under SLOCC \cite{DVC}. Thus,
answering if a state is more entangled than other will depend on
what criterion one is adopting.

Consider an $m$-partite state with Hilbert space
$H=\bigotimes_{i=1}^{m}H_i$. We call $P_{k}^{m}=\{ A_j \}_{j=1}^N$
a $k$-partition of $\lbrace1,2,...,m\rbrace$ if: \textbf{1.} $A_j
\subset \{1,2,...,m\}$; \textbf{2.} $A_i \cap A_j = \emptyset,
\forall \hspace{0.1 cm} i \neq j$; \textbf{3.} $\bigcup_{i} A_i =
\lbrace1,2,...,m\rbrace$; \textbf{4.} $\sharp A_j \leq k$. The
number $k$ is called the diameter of the partition $P_{k}^m$. The
set of all $k$-partitions of $\{1,2,...,m\}$ will be denoted by
${\cal P}_{k}^m$. With this concept, one can define
factorizability and separability subjected to a partition, and
also subjected only to the diameter of the partitions.
\begin{definition}
We say that a state $\rho$ is $P_{k}^m$-factorizable if, for a
fixed $P_{k}^m$, it can be written as
$\rho=\rho_{A_1}\otimes...\otimes\rho_{A_n}$, where $\rho_{A_j}$
is a density operator on $H_{A_j}=\bigotimes_{i \in A_j}H_i$. A
state is $P_k^m$-separable if it can be written as a convex
combination of $P_{k}^{m}$-factorizable states. Finally, we call
$k$-separable a state $\rho$ which can be written as a convex
combination of $P_{k}^{m}$-factorizable states , where $P_{k}^{m}$
may vary for each pure state.
\end{definition}
%Note that for $k=1$, these are the usual notions of factorizable
%and separable states. A 2-separable state is allowed to have pair
%entanglement (in any of its pairs), although no three particle
%entanglement can exist. With this definition every $m$-partite
%state $\rho \in H$ is trivially $m$-separable.

Let us denote $S_k(H)$ the set of $k$-separable states on $H$.
Clearly they form a chain $S_1(H) \subset S_2 (H)
\subset...\subset S_{m-1}(H) \subset S_m(H) = D(H)$, where $D(H)$
denotes the set of density operators on $H$. As each of these sets is
closed and convex, there exists an Hermitian operator $W$ such that tr$(W\rho) <
0$, and tr$(W\sigma) \geq 0$ $\forall$ $\sigma \in S_k(H)$
\cite{HHH2}. One call such $W$ a $k$-entanglement witness.
%Whenever $\rho$ has some kind of entanglement, there is some $W$ which witnessed it. However, there is no ``universal entanglement witness'': given $\rho$ and a
%specific kind of entanglement one has to search for its witnesses.

Although several entanglement monotones applicable to multipartite
states are known \cite{CKW, VP, BV1, EAP, PVDC}, only two
approaches, up to now, can be applied to the quantification of the
different kinds of multipartite entanglement discussed above
\footnote{The other measures are either based on bipartite
entanglement concepts, such as the localizable entanglement, or can
only distinguish entangled from fully-separable states.}: the
relative entropy of entanglement \cite{VP} and its related measures
\cite{VP, EAP} and the witnessed entanglement \cite{BV1}. The first,
of great importance in the bipartite scenario, is based on the
minimization of some distance between the state under question and
the sets $S_k(H)$. The second, recently introduced in Ref.
\cite{BV1}, includes several well studied bipartite and multipartite
entanglement measures and quantifies entanglement based on the
concept of optimal entanglement witnesses. In this paper, due to
some particular properties, such as the linearity of the objective
function, we will consider the witnessed entanglement:

\begin{definition}For a $m$-partite state $\rho \in D(H)$, its witnessed
$k$-partite entanglement is given by
\begin{equation}\label{Ew}
E_{W}^{k}(\rho)=\max \{0, -\min_{W\in\mathcal{M}}Tr(W\rho)\},
\end{equation}
where $\mathcal{M}=\mathcal{W}_{k}\cap\mathcal{C}$,
$\mathcal{W}_{k}$ is the set of $k$-entanglement witnesses and
$\mathcal{C}$ is some set such that $\mathcal{M}$ is compact.
\end{definition}

Having this definition in mind we can see what are the restrictions
imposed
by the existence of a MME-state $\rho$ on its optimal entanglement
witness $W$.

Let $\rho = \sum_{j}q_{j}\ket{\phi_{j}}\bra{\phi_{j}}$ be the spectral
decomposition
of $\rho$, and $\{ p_i, \ket{\psi_i}\}$ another ensemble describing it.
Then, $\ket{\psi_i}=\sum_j c_{ij} \ket{\phi_j}$, with coefficients $c_{ij} = u_{ij}\sqrt{\frac{q_j}{p_i}}$. In the case
where $\rho$ is maximally entangled with
entanglement $E$, $W$ must be optimal for every
$\ket{\psi_{i}}$ and $\ket{\phi_{j}}$, which allows us to write
for one specific element $\ket{\psi_{k}}$,
\begin{eqnarray}
-E &=& Tr(W\ket{\psi_{k}}\bra{\psi_{k}}) \nonumber \\
   &=& \sum_i |c_{ki}|^{2}\bra{\phi_{i}}W\ket{\phi_{i}} + \sum_{i
\neq j} c_{ki}^{\ast}c_{kj}\bra{\phi_{i}}W\ket{\phi_{j}} \nonumber
\\ &=& -E + \sum_{i\neq j}
c_{ki}^{\ast}c_{kj}\bra{\phi_{i}}W\ket{\phi_{j}},
\end{eqnarray}
which implies
\begin{equation*}
\sum_{i \neq j} c_{ki}^{\ast}c_{kj}\bra{\phi_i}W\ket{\phi_j} = 0.
\end{equation*}
As this equality must be true for every ensemble describing
$\rho$, $\bra{\phi_i}W\ket{\phi_j} = 0$ and $W$ is proportional to
the identity matrix in the support of $\rho$, with $-E$ as the
proportionality constant. Being $E$ the highest value of
entanglement and, therefore, the modulus of the lowest eigenvalue
possible among all entanglement witnesses, each eigenvector
$\ket{\phi_j}$ of $\rho$ is an eigenvector of $W$ too. Thus $W$
can be written as
\begin{equation}\label{form}
W = \underbrace{(-E)I}_{Supp(\rho)}\oplus
\underbrace{D}_{Supp^{\bot}(\rho)},
\end{equation}
where $D$ is some matrix such that the constraints imposed by
$\mathcal{C}$
are satisfied. Here again this demanding condition is not sufficient to
rule
out the existence of mixed maximally entangled states. As a
counterexample,
consider the best separable approximation measure \cite{KL}, $BSA^{k}(\rho) = 1 - \lambda$, where $\lambda$
is the optimal value of the following optimization problem:
\begin{equation}
\max \lambda, \hspace{0.2 cm} s.t. \hspace{0.2cm}
\rho = \lambda \sigma + (1 - \lambda) \pi,
\end{equation}
with $\sigma \in S_{k}(H)$, $\pi \in D(H)$ and $\lambda \in [0,
1]$. It can be written alternatively as Eq. \eqref{Ew} with ${\cal
C} = {\cal f}W \hspace{0.1 cm}|\hspace{0.1 cm} W \geq - I{\cal g}$
\cite{BV1}. For the following family of mixed states
\begin{equation}\label{WGHZ}
\rho_{q} = q\ket{W}\bra{W} + (1 - q)\ket{GHZ}\bra{GHZ},
\end{equation}
$BSA$ was calculated in Ref. \cite{BV1}, using the numerical method
presented in Ref. \cite{BV2}, and shown to be composed only of
maximally entangled states, with respect either to 1 and
2-entanglement. Note, nonetheless, that despite $BSA(\rho)$ being
an entanglement monotone \cite{KL}, it is a quite odd quantity, as
every entangled pure state is maximally entangled.

An important measure of multipartite entanglement is the generalized
robustness of entanglement \cite{S},
\begin{equation}
R^{k}(\rho) = \min s, \hspace{0.2 cm} s.t. \hspace{0.2 cm} \frac{1}{1 +
s}\rho
+ \frac{s}{1 + s}\pi = \sigma,
\end{equation}
where $\sigma \in S_{k}(H)$ and $\pi \in D(H)$. It gives good
bounds for the maximum fidelity of teleportation, the distillable
entanglement, and the entanglement of formation \cite{BV1}, and has
important applications in the study of threshold of errors in
quantum gates \cite{HN}. In Ref. \cite{BV1}, it was shown that $R$
can be written as Eq. \eqref{Ew}, with ${\cal C} = {\cal f}W
\hspace{0.1 cm}|\hspace{0.1 cm} W \leq I {\cal g}$.

\begin{lemma}
For every state $\rho \in D(H)$,
\begin{equation}\label{rob}
\max_{\sigma \in S_{k}(H)}tr(\rho \sigma) \geq
\frac{tr(\rho^{2})}{1 + R^{k}(\rho)}.
\end{equation}
\end{lemma}
\begin{proof}
From the theory of convex optimization and Lagrange duality
\cite{BV}, the optimal value of the L.H.S. of Eq. \eqref{rob} is
given by the solution of the following convex problem
\begin{equation}\label{dual}
\min \lambda \hspace{0.2 cm} s.t. \hspace{0.1 cm} \lambda I - \rho
\in {\cal W}_{k}.
\end{equation}
Let $W = \lambda_{opt} I - \rho$ be an optimal solution of
\eqref{dual}. Since, $W/\lambda_{opt} \leq I$, we find that
$R^{k}(\rho) \geq tr(\rho^{2})/\lambda - 1$, from which the result
follows.
\end{proof}

\begin{theorem}
There do not exist, for any $k$, mixed maximally $k$-entangled
states according to the generalized robustness of entanglement.
\end{theorem}
\begin{proof}
We will prove by contradiction that it does not exist an optimal
entanglement witness of the form \eqref{form}. Assume that $\rho'$
is a mixed maximally entangled state with spectral decomposition
$\rho' = \sum_{i=1}^{m}\lambda_{i}\ket{i}\bra{i}$. Then, by the
dual definition of $R$, it is easily seen that $\rho = \lambda
\ket{1}\bra{1} + (1 - \lambda)\ket{2}\bra{2}$ is also maximal
entangled for every $\lambda \in [0, 1]$. Thus we consider,
without loss of generality, states of rank two. For the
generalized robustness, the matrix $D$ of Eq. \eqref{form} must
satisfy $D \leq I$. Hence, since for every entanglement witness $W
= (-E)I \oplus D$, with $D \leq I$, $W' = (-E)I \oplus I$
is another witness as optimal as $W$ for $\rho$, we may assume
throughout this canonical form. Letting $P$ represents the
projector onto the support of $\rho$, $W = I - (1 + E)P$. By
assumption, $R(P) = 2E$. From Lemma 1, we find that for some
$\sigma \in S_{k}(H)$:
\begin{equation}
tr(P\sigma) \geq \frac{tr(P^{2})}{1 + R(P)} = \frac{2}{1 + 2E}.
\end{equation}
Therefore,
\begin{equation}
Tr(W\sigma) = 1 - (1 + E)tr(\sigma P) \leq 1 - \frac{2(1 + E)}{1 + 2E}
< 0,
\end{equation}
which contradicts the fact that $W$ is a $k$-entanglement witness.
\end{proof}

In Ref. \cite{BV1}, a family $E_{m:n}$ of infinite entanglement
monotones which interpolates between the best separable
approximation measure and the generalized robustness was proposed.
For  fixed $m$ and $n$, $E_{m:n}$ is given by Eq. \eqref{Ew}, with
${\cal C} = {\cal f}W \hspace{0.1 cm}|\hspace{0.1 cm} -mI \leq W
\leq nI{\cal g}$. They provide a tool to the observation of the
(smoothly) transition between the regime where there exist mixed
maximally entangled states and the regime where all MMES are pure.
Fig. 1 shows the transition for the family of states given by Eq.
\eqref{WGHZ}.
\begin{figure}
\begin{center}
\includegraphics[scale=0.35]{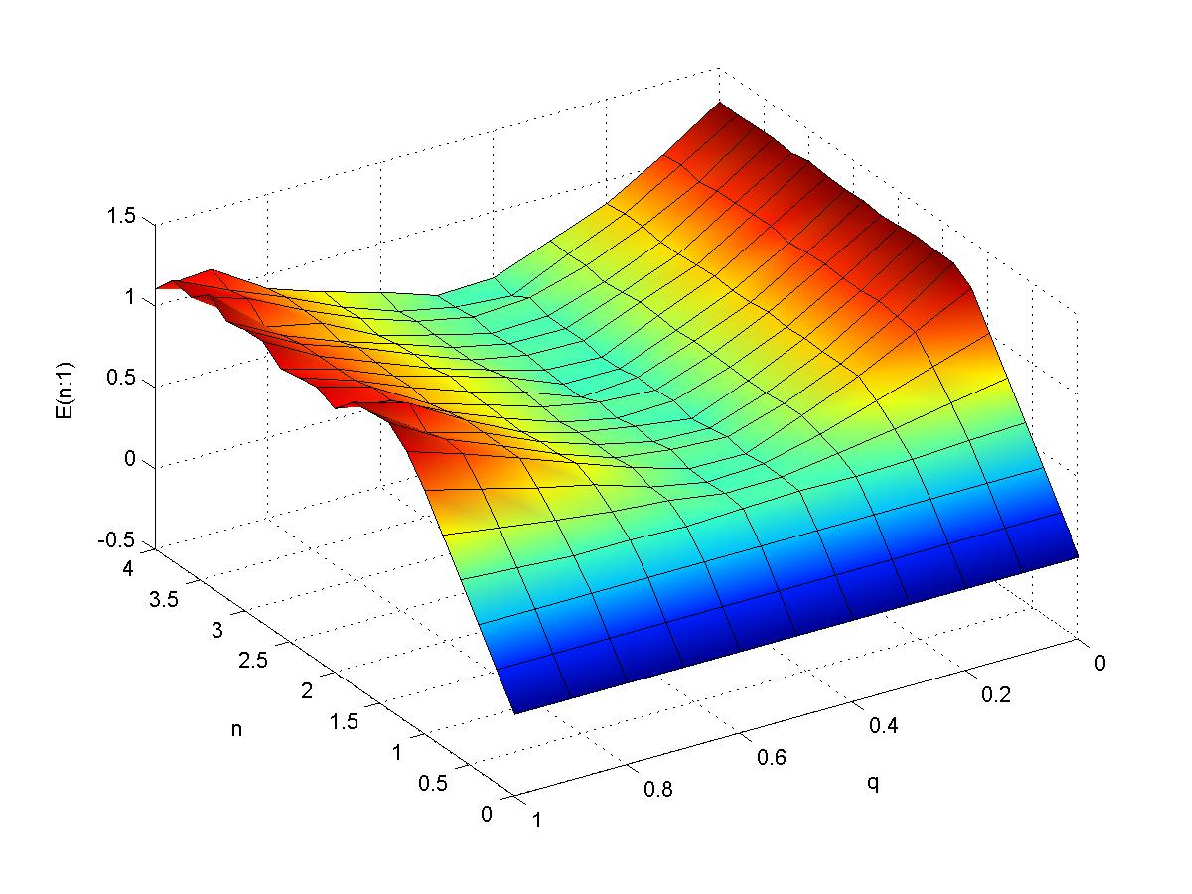}
\caption{(Color online) $E^{1}_{n:1}(\rho_{q})$ for $0 \leq n \leq
4$ and $0 \leq q \leq 1$. When $n\ll1$,
$E^{1}_{n:1}(\rho)=nBSA(\rho)$ and MME-states are possible. In the
other limit $E^{1}_{n:1}(\rho)=R^1(\rho)$ \cite{BV1}, and all
maximally entangled states are pure. The irregularities in the
figure are due to the approximative nature of the numerical method
used \cite{BV2}.}
\end{center}
\end{figure}

In conclusion, we have analyzed the existence of maximally mixed
entangled states in the bipartite and multipartite scenarios. In the
first, we showed that, although monotonicity under LOCC do not
exclude MMES, partial additivity and asymptotic continuity together
with monotonicity do. We then extended this result
%introducing the concept of \textit{faithful polygamy of
%entanglement}. The term \textit{polygamy} comes from the possibility
%of $k$-entanglement, for every $k$, between an arbitrary number $n$
%of parties, whereas \textit{faithful} represents the constrained
%nature of entanglement even in this case, since for every $k$ and
%$n$ chosen, maximally entangled states must be pure, not being
%correlated with anyone else.
to multipartite systems by showing that maximally entangled
(multiparticle-)states are pure. It is now time to ask: what are the
physical consequences of this result? One can easily note that every
pure state must be completely uncorrelated with any other system (if
not it should be written as a non-trivial convex combination, what
characterizes mixed states). This notion gives us a solid background
to propose the \emph{faithful polygamy of entanglement}, which
states that all maximally entangled states are (classically and
quantically) uncorrelated with any other system. One can even
propose this condition as another requisite for a good multipartite
entanglement quantifier. Furthermore, it is also important to stress
that this polygamy holds for all kinds of entanglement, it is, every
time the system reaches a maximum amount of entanglement according
to any partition, it becomes ``free'' of its environment.

DC and FGSLB acknowledge financial support from CNPq. The authors
thank M.F. Santos for useful discussions.

\end{document}